# Using the DELPHI Method for Model for Role Assignment in the Software Industry

Daniel Varona  
CulturePlex Laboratory  
Western University  
London, Canada  
dvaronac@uwo.ca

Luiz Fernando Capretz  
Dept. of Electrical & Computer Engineering  
Western University  
London, Canada  
lcapretz@uwo.ca

*Abstract*—Over the past two decades, there has been a growing interest in modeling the elements that need to be considered when assigning people to roles in software projects, as evidenced by the number of available publications related to the topic. However, for the most part, these studies, have taken only a partial approach to the issue. Some have focused on the target role´s competency profile, while others have tried to understand the preferences of software developers for activities linked to certain roles and the relationship between these preferences and the candidate´s personal traits, to mention only two examples. Our research aims to find elements that can be integrated into an allocation model that complements current approaches by including competencies, personal traits, and project team building theories. To do so we performed an expert's consultation exercise using the DELPHI method; which allowed us to validate a set of patterns related to different candidate´s personal traits, and the link between the team´s motivational motors and their roles within the software development.

## I. INTRODUCTION

The literature we reviewed indicates that the models used for role assignment contain multiple insufficiencies in the descriptions of the evidence used to support the assignments made by certain project teams [1-10]. This, coupled with the need to contextualize a role-to-role mapping model in the software industry environment, supports the need to use expert consultation methods to design a proposed model for assigning people to roles on software projects.

In this research it was decided to use the Delphi method as a method of expert consultation. The Delphi method is a method that stands out for its wide application in dissimilar fields [11], and that has been used, for example, in Cuba, in technologies development related domains [6, 12-15], and in Canada, in cases such as [16, 17], to name just a few examples. Its versatility and ability to adapt to different contexts allows us to take advantage of the benefits of group debate, while preserving anonymity by using feedback flows through several channels, and permitting asynchronous participation when participants are geographically dispersed or subject to commitments that impede their ability to participate in face-to-face meetings, as was the case in this investigation.

Our main goal in this study was to identify those elements that can be integrated into a model for role assignment in software development projects. Compared to current models, which contain only partial approaches to the role assignment issue, we seek to design a model integrating a set of tools aimed at complementing the current approaches, when integrated into a single mechanism that is focused on the assessment of skills, personal traits and the candidates´ contributions to the team, based on their individual traits.

## II. METHOD

The Delphi method was applied in several rounds, pursuing different objectives during the investigation, as can be seen in Fig. 1. In the first stage, the method was used to identify the criteria for the selection of experts who would participate in the next stages. Those selection criteria were used to find a new set of experts—selected in further stages of the method— who helped to identify and validate the elements to be taken into account in the development of a model role assignment on software project teams.

The second stage was developed in two rounds aimed at identifying which elements should be considered in role assignment in software development projects. In this stage (the second round), the elements resulting from the first round were validated. The Kendall concordance coefficient was used to determine consensus among participants at each stage of implementation of the method. 60% was set as the reference value for the concordance threshold, as it is a generally accepted value among the reviewed investigations.

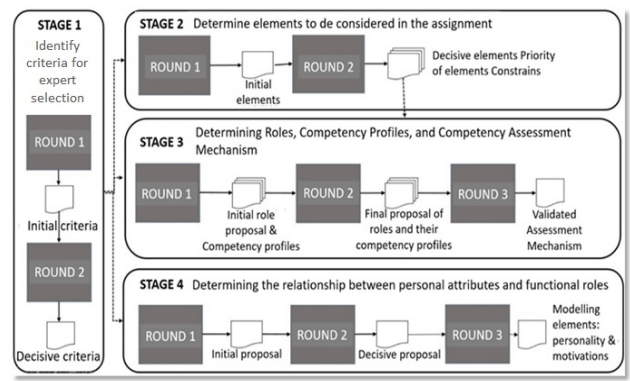

Fig. 1. Representation of the application of the Delphi method.

Stages three and four were developed in three rounds each. The third stage of the method was necessary to determine not only the main roles to be assigned in software projects, and their competency profiles, but also to validate a proposed mechanism for assessing the competence of potential candidates for certain roles.

Finally, the fourth stage connected the relationship between the functional roles of software development and certain personal attributes of role-runners. The Results and Discussion section discuss the method of execution in more detail.

### III. RESULTS AND DISCUSSION

As mentioned in the Methods section, this section details the particularities of each of the stages of execution of the Delphi consultation method through the analyses of each round´s outcomes. The stage one focusses on identifying the elements that future experts must comply with, and its presented next.

*A. Stage 1: Expert Selection*

The literature reviewed highlights several ways to make a correct selection of the experts who will staff the Delphi method execution group, among which the assessment and analysis of the candidate's curricular synthesis stands out. On the one hand, there is a tendency to assess an expert's competence according to elements such as scientific degrees, academic qualifications, teaching and research, managerial positions in which the candidate has performed, as well as the candidate´s scientific production (expressed in: publications, patents, intellectual property records, and conference participation). However, these elements do not always constitute a guarantee of the expertise of a given candidate. Specifically, in this investigation, it was necessary that the experts we consulted had practical experience. It can then be noted that, generally, successful specialists in software management and software development have neither a scientific degree, nor experience in teaching or scientific research. It was then decided to use the curriculum synthesis analysis for the selection of experts, based on a set of criteria defined for the purposes of this research, while applying the Delphi method.

As can be seen in Fig. 1, the selection of experts took place in the first stage of development of the method; executed in two rounds, where 11 professionals from five Cuban institutions, with at least 10 years of experience, participated in the development and management of software projects contracted with national and international clients. In the first round, participants were individually asked to provide criteria in order to be considered in qualifying as an expert in software developer/management for the purposes of the present research. After a debugging process, where repetitions or similarities were removed, a general list was produced that became the selection criteria matrix for participants in further stages of the Delphi consultation method.

In the second round the objective was to expose the selection criteria matrix to all participants and to identify those criteria in which the level of concordance between the participants exceeded 60%. The final selection criteria that resulted from this round for the selection of experts is listed below:

- More than 10 years of experience actively and successfully participating in software development.
- Knowledge of software engineering, and management techniques and contracting management in software projects.
- To have participated in and published research on experiences related to human resources management in software projects.
- Having successfully led teams with a minimum of 3 members in at least two medium or high complexity level projects[1].

The participation of the experts surveyed in this stage was completely anonymous through the web tool for survey processing Survey. These selection criteria were used for selecting the experts who participated in the rest of the stages. We gathered 15 experts willing to contribute to the study. Below can be found a set of features that guarantee the quality of the selected experts:

- 100% of the chosen experts come from the Cuban software industry.
- 100% of the selected experts know and have applied at least four software development methodologies.
- 100% of experts selected have held various roles, such as Project Manager, Analyst, Designer, Programmer, Tester, and roles related to quality management, maintenance and support tasks.
- 100% of the selected experts have experience developing software projects in different domains of application.
- More than 70% of experts selected have participated in the process of assigning people to roles in various projects of different types, categories, and complexity.
- More than 50% of selected experts have been involved in the development of between five and seven software development projects.
- 100% of selected experts have been linked to research related to human resources management in software projects and have published in journals and/or conferences related to the industry.

The main results obtained from each of the remaining stages of the execution of the Delphi method are described below.

*B. Stage 2: Determining the elements to be considered in role assignment on software project teams*

Graphically represented in Fig. 1, the second stage of the Delphi method application is aimed at identifying the elements to be considered when assigning roles on software project teams. This stage was developed in two rounds, with the consultation of fifteen experts. Once they had kindly confirmed their acceptance, each of these experts was sent specific and specialized literature related to the present study.

---

[1] Experts believe that it is necessary, to evaluate the complexity of a software project, to analyze – among other things – elements such as: importance, size, team composition (size, roles, experience), time available for its development, and technology used

In the first round, ten of the participant experts were surveyed, using open-ended questions, with the intention of performing an initial identification of the elements to be considered during role assignment on software development teams. For the second round the survey was redesigned with semi-closed questions based on the results obtained in the first round; and it was applied to all 15 participant experts.

Drawing on the survey results, where there was a minimum concordance of 60%, the resulting elements to be considered during role assignment in software development teams were:

- The candidate´s competence expressed in the areas of theoretical domain, techniques, and methods related to the work that will be assigned.
- The candidate's experience in similar projects and in the target role.
- The candidate´s personal attributes favoring the synergy of the project team, to be determined after the assignment.
- The Candidate´s availability in relation to the development plan of the organization's project portfolio.
- Size of the work team to which the candidate will be appointed after assignment.
- Project data: complexity, importance and risk.

The following element did not reach 60% concordance among experts:

- Geographical location of the candidate (It should be noted that this element reached values greater than 50% concordance among experts). The remainder of experts thought this element was only valid on projects taken on by geographically dispersed teams or when the organization exhibits a cloud-based open development model.

There was, among the experts, agreement on the relevance of considering these elements and on prioritizing them. The three elements that obtained the highest priority values were: maximizing a candidate´s competence, maximizing team synergy, and minimizing the team size, in that order.

In addition, in analyzing the consensus that might exist in relation to the influence of combining various elements and specific roles, there was agreement on the following points:

- An employee can perform multiple roles on the same project, as well as be linked to multiple projects.
- Although an employee serving as a project leader, can have multiple roles on the same project, he/she can only work as a project leader on another project(s), if the project(s) has a common application domain.
- The decision to assign employees to other projects or another role within the same project will depend on their workload.

There was no agreement among the experts regarding the following:

- The existence of roles that always need maximum competence from candidates in general, or under specific conditions, or what these conditions would be.
- The existence of roles that demand greater synergy with the rest of the project team members, or what these roles would be.

However, before applying a role assignment model that takes into account the individual competencies of candidates, it is necessary for organizations to have formalized the roles needed for their projects and the skills required from candidates for proper performance in these roles. Among the related specialized software development literature the following role proposals can be found: role competencies from RUP and XP software development methodologies [18] [19] (the most commonly mentioned), the invariant roles for software projects proposed in [15], and the professional definitions provided by [20].

In the particular case of this study, the author points out the need to find a balance with the specialization proposed by Andre in [15]. This vision integrates the theoretical perspective provided by the software development methodologies mentioned above, and the analysis of practical experience of a set of experts; as well as the generalization that exhibits the definition of roles present in [20]. We wanted to ask what it was that supports the idea of a link between a Cuban context modeled by Andre, in her thesis [15], and the definitions supporting the global understanding of researchers currently investigating the relationship between the personal attributes and roles of software development.

To do this, it was necessary to develop a third stage of the Delphi method focused on validating roles, and creating competency profiles and a competence assessment mechanism.

*C. Stage 3: Role Determination, Competence Profiles, and the Competence Assessment Mechanism*

The third stage of the method was divided into three rounds, in which the 15 experts, who agreed to collaborate with the study and whose characterization has already been outlined above, participated. The goal in this stage is to identify software development roles, their competence profiles, and validate a mechanism for assessing these competences in potential candidates.

It is worth noting that during round number one we sent the Andre´s invariant roles, the CPP´s role definitions, and asked the participants to for classify those roles into the following general roles of software development: Project leader, Analyst, Designer, Programmer, Tester, and Roles associated with maintenance activities. On their response, each general role had to be accompanied by their competence profile.

Table I represents the description of the query results. After removing duplications, the authors have inserted experts' suggestions into the columns "Invariant roles proposed by Andre and Roles defined by the CPP" and

"Comments on competences"[2]. The resulting information is used for the initial proposal for competences for each of the general roles that will be validated in further rounds.

TABLE I. GENERAL ROLES CLASIFICATION RESULTING FROM ROUND ONE

| General roles | Invariant roles proposed by Andre and Roles defined by the CPP | Comments on competences |
|---|---|---|
| Project leader | Invariant Role/CPP role definition 1 .. Invariant Role/CPP role definition N | Generic/Technical competence 1 . Generic/Technical competence N |
| Analyst | Invariant Role/CPP role definition 1 .. Invariant Role/CPP role definition N | Generic/Technical competence 1 . Generic/Technical competence N |
| Designer | Invariant Role/CPP role definition 1 .. Invariant Role/CPP role definition N | Generic/Technical competence 1 . Generic/Technical competence N |
| Programmer | Invariant Role/CPP role definition 1 .. Invariant Role/CPP role definition N | Generic/Technical competence 1 . Generic/Technical competence N |
| Tester | Invariant Role/CPP role definition 1 .. Invariant Role/CPP role definition N | Generic/Technical competence 1 . Generic/Technical competence N |
| Role associated with maintenance activities | Invariant Role/CPP role definition 1 .. Invariant Role/CPP role definition N | Generic/Technical competence 1 . Generic/Technical competence N |

In round number two, the authors took the liberty of modifying the classic consultation method and invited five individuals with at least ten years of experience in each of the general roles investigated to participate, making a total of 35 guests. A workshop was organized in six different locations (one for the analysis of each role) where at least two of the experts participated as facilitators along with the five guests corresponding to a given role. Before the start of the exercise, participants were asked not to exchange ideas, unless it was necessary for clarification purposes.

Each group was given a list of general roles comprising a more specific set of roles –with the information shown in Table I- related to software development, and a survey consisting of two semi-open questions. The first question digged whereas they accepted the provided the classification and requested an explanation in case negative; and the second, requested a mark with an "X", from the competences that constituted the general role competence pro those with which they agreed should integrate the profile of the role under analysis. On the second question, experts could propose changes.

---

[2] Nor competences or comments are listed for space reasons; and because each organization its different and must determine their roles, define their role´s competence profile.

Below are those elements where experts reached concordance higher than 60%:

- 100% of the consulted experts were in favor of generalizing the roles, even though they were not asked about it. The author finds it interesting that in the judgment of the experts, specializing people into particular activities conditions them to be resistant to changes in methodology, and in addition, that excessive division of responsibilities conditions the amount of overlap between the roles.

- 100% of the consulted experts agreed that the competences covered by the role of Analyst must include CPP's descriptions for business analysts, system analysts, and data analysts.

- More than 88% of the consulted experts agreed that the UI Designer role does not necessarily have to be sub-classified into the role of Designer, but instead has more to do with the Role of Analyst.

- More than 72% of the consulted experts agreed that the competences associated with configuration management, change management, security management, and documentation management related activities must be present in all roles, and not only with the role of Project Leader for the first two, with the role of Designer for the third, or with the role of Tester for the last case.

On the other hand, there was a trend that, although it did not exceed 60% of the concordance among the experts the author's estimate it is important to mention as it did reach 56% of concordance. This is displayed below:

- The role of User Interface Designer must be: either, a) performed by an information architecture specialist or b) by a trained individual in Graphic Design, or c) be a subcontracted service with an analyst proposed model as a starting point. This would give the product an added value based on aesthetics, and message effectiveness by embedding tools that are unfamiliar to the software developer.

It should be noted that there was no concordance on the next element:

- Which General Role to sub-classify the invariant role of Technical Specialist from Andre´s proposal in, based on its description.

Finally, a competence assessment mechanism was designed [21] consisting of three exercises: a group dynamic, a questionnaire, and an interview. The exercises were aimed at assessing candidates´ ability to fulfill the skills necessary to cover the role to which he/she aspires. In a third round of consultation on the validity of the proposed mechanism, the mechanism was mailed to experts, along with a questionnaire aimed at dig.

The questionnaire, designed for the validation of the competence assessment mechanism, consists of 7 reflective and open-ended questions, and concludes with a space where experts can give their personal opinion, which is very important to understand the suggestions on the proposed mechanism.

Reflection questions allow us to record certain aspects in order to facilitate the processing of the results yielded by the questionnaire. The scale used is framed by values between "One," representing the lowest level, and "five," representing the highest level. Yet, it was necessary to use equivalence balance, as quantitative and qualitative assessments are made in the questionnaire. Table II shows a set of evaluation criteria balancing equivalences in order to make a general analysis that provides an integrative result on the responses of the query.

TABLE II.  EQUIVALENCE BALANCE FOR QUESTIONNAIRE PROCESSING

| Qualitative criteria | Criterion expressed in % | Numerical criterion |
|---|---|---|
| Very high or there is a very remarkable contribution | Greater than 90% | 5 |
| High or There is a notable contribution | Greater than 75% | 4 |
| Average or there is an average contribution | Greater than 50% | 3 |
| Low or there is a slight contribution | Greater than 25% | 2 |
| Null or there is no contribution | Less than 25% | 1 |

Table III is a sample of the consultation method processing sheet for the validation of the proposed mechanism for competence assessment.

TABLE III.  PROCESSING SHEET FOR PROCESSING CONSULTATION RESULTS

| Criterion | PPI | KAC | RAC | EPIC | PC | AP | PEG | TOTAL |
|---|---|---|---|---|---|---|---|---|
| Expert 1 | | | | | | | | |
| .. | | | | | | | | |
| Expert 15 | | | | | | | | |
| Sum Frj | | | | | | | | |
| Average Score | | | | | | | | |
| Maximum Score | 5 | 5 | 5 | 5 | 5 | 5 | 5 | |
| Acceptance Index | | | | | | | | |

The columns "PPI", "KAC", "RAC", "EPIC", "PC", "PA", and "PGE" correspond to: the Proposal´s Perceived Importance, Knowledge Acquisition Contribution, Realization Assessment Contribution, Evaluation Process Improvement Contribution, Proposal´s Content, Proposal Applicability, and Proposal´s General Evaluation respectively.

To determine the concordance of criteria among the experts in this third round, the Hypothesis Significance Test was carried out. For that Kendall's concordance coefficient (W) was calculated, supported by the Statistical Product and Service Solutions (SPSS) statistical software. Figures 2 and 3 show test results. The results obtained from the applied questionnaire constitute the entry to the calculation of W. Kendall coefficient "W" values should range between zero and one ($0 < W < 1$), if W reaches a value of one ($W = 1$) then a total concordance exists; the higher the value of W, i.e. the closer it gets to one, the greater the concordance among the experts. Concordance is considered acceptable if $W \geq 0.5$.

| | N | Mean | Std. Deviation | Minimum | Maximum | Percentiles | | |
|---|---|---|---|---|---|---|---|---|
| | | | | | | 25th | 50th (Median) | 75th |
| IP | 15 | 5.0000 | .00000 | 5.00 | 5.00 | 5.0000 | 5.0000 | 5.0000 |
| COC | 15 | 5.0000 | .00000 | 5.00 | 5.00 | 5.0000 | 5.0000 | 5.0000 |
| CRV | 15 | 5.0000 | .00000 | 5.00 | 5.00 | 5.0000 | 5.0000 | 5.0000 |
| CMPE | 15 | 4.5333 | .51640 | 4.00 | 5.00 | 4.0000 | 5.0000 | 5.0000 |
| CP | 15 | 4.5333 | .51640 | 4.00 | 5.00 | 4.0000 | 5.0000 | 5.0000 |
| AP | 15 | 3.5333 | .51640 | 3.00 | 4.00 | 3.0000 | 4.0000 | 4.0000 |
| EGP | 15 | 4.4000 | .50709 | 4.00 | 5.00 | 4.0000 | 4.0000 | 5.0000 |

Fig. 2.  Screenshot of the description of the evaluation in SPSS

| Ranks | Mean Rank |
|---|---|
| IP | 5.27 |
| COC | 5.27 |
| CRV | 5.27 |
| CMPE | 3.77 |
| CP | 3.77 |
| AP | 1.27 |
| EGP | 3.40 |

| Test Statistics | |
|---|---|
| N | 15 |
| Kendall's W<sup>a</sup> | .688 |
| Chi-Square | 61.932 |
| df | 6 |
| Asymp. Sig. | .000 |

a. Kendall's Coefficient of Concordance

Fig. 3.  Screenshot of the summary of evaluation and result of Kendall Coefficient (W) in SPSS

The null (Ho) and alternative hypotheses (H1) are declared as follows:

Ho: There is no concordance among the experts. $W < 0.5$

H1: There is agreement among the experts. $W > 0.5$

As can be seen in Fig.3, the value of W (0.688) is higher than 0.5, so the concordance among experts is considered acceptable.

Note that in the same figure, the value of significance (Asymp Sig. = 0.000) is less than 0.001, so the null hypothesis can be rejected (Ho), and it can be declared, with high statistical significance that there is agreement among experts on the validity of the competence assessment mechanism.

Just as it was necessary to formalize what was related to the edge of competence, it was also necessary to identify those elements that, from an individual perspective, deserved interest in favor of the synergy of the team. The characteristics of the implementation of the fourth stage of development of the method are described below.

*D. Stage 4: Determining the relationship between personal traits and functional roles*

Fifteen experts participated in this consultation stage. The stage was divided into three rounds and communication with the participants was maintained via email, in such a way that the experts' criteria were not compromised in a debate.

Before asking about those personal attributes that, in the expert´s opinion, favored the synergy of a team, it was noted that for the purposes of the study, Andre's definition in [15] is assumed, in conjunction with the following synergy of the team definition: "Collective action and creation; unification, cooperation and cause contest to achieve results and joint benefits; concerted action towards common objectives".

As a result of the debugging of the proposals made by the experts; and after eliminating duplicate opinions, and widespread those seemingly different issues, but with a common axis; result an initial proposal to be re-analyzed to assess the overall consensus. In the second round, 60% or more of the opinions find consensus on the point that in order to promote the synergy of the team, the following elements should be taken into consideration during the candidates' analysis, in order of priority:

- Personality traits or personality types, or character analysis displayed by candidates in their contribution to the team.
- Belbin roles distribution among team members.
- McLelland Motivations distribution among team members.
- Interpersonal relationships among team members in relation to conflict management.

It is also in the authors' interest to show three elements in which the consensus exceeded the 50% concordance among experts:

- Ability to resolve conflicts
- Ability to generate conflict
- Team size

In contrast, no agreement was reached on whether the analysis of Belbin's roles distribution should be directed to form homogeneous or heterogeneous teams, which the authors find interesting. It should be clarified that in [15] it is recommended that this topic be delved into more deeply.

To model Belbin roles distribution within the team, we presented the experts with the patterns defined by Andre in [15]. However, a third round of consultation was needed to validate personality traits, and McLelland motivations related patterns.

*1) Analysis of instruments that measure traits, personality types, and the character exhibited by software engineers*

As a reference we used a previous study on the MBTI personality type of software developers and their preference for certain software development tasks. The main results of the experiment can be found in [22] related to the preferences of software developers for certain tasks within the roles under study; and in [23] related to the identified patterns of the latter.

*2) McLelland Motivations*

Similarly, we used a previous study on the motivational engines of 100 software developers available in [24]. In both cases, concordance values greater than 60% approval among experts were achieved, validating the results from both studies as instruments complementing role assignments in software development, and that correspond to their assessments of the second round in the second stage of the method. The MBTI patterns described in [23] received a concordance value of 82%, and the McLelland patterns described in [24] reached a concordance value of 79% among the experts.

## IV. Conclusions

The design of the DELPHI consultation method was executed in 4 stages and organized in a total of 10 rounds. 60% was chosen as a benchmark for consensus determination among experts in the assessment of the Kendall coefficient. A total of 26 experts participated in the exercise, eleven in the preparatory stage (Stage 1) and another fifteen in the functional stages (Stages 2-4). In addition, a total of 35 developers with at least 10 years of experience in one of the roles they consulted on, participated as facilitators in one of the rounds.

Among the experts who participated in the 3 functional stages of the method are those with the following credentials: experience in at least 4 different methodologies of software development, experience performing in various software development roles in projects of varying complexity and in different application domains; and have shown evidence of active participation in software project management related research.

The elements identified for assigning roles in software development projects integrate current hiring model trends, which include: 1) maximizing the skills required by role competence profiles, 2) maximizing team synergy; which also frames the candidate's contribution to the team, according to Belbin's roles, 3) the candidate's natural disposition analysis for the role, given the Myers-Briggs type indicators, 4) the candidate´s primary and secondary motivations according to McClelland's theories; and 4) minimizing the size of the project team, in that order.

The experts pointed to the need for the hiring agents to determine a priority rating of the elements that have been proposed, to bring them closer to the specific needs of the company at the time of the allocation.

In certain contexts, and taking into consideration minimizing the size of the project team, the experts suggest a set of elements that can help to standardize the assignment of the same employee to multiple roles inside and outside the same software development project.

Although the experts were not asked, they agreed that a balance must be found between competence, skills, and a candidate´s personal traits at the time of the assignment. This might have a positive impact on future resistance to changes in technology and methodology, conflicts related to the structure of project teams, and other conflicts that could manifest among highly specialized candidates without the minimum capacity to adapt to new or changing working environments, such as software projects.

The team's synergy analysis assesses an individual´s natural disposition according to their Myers-Brigs type indicator, individual and team competences, individual contribution of Belbin's roles to the team, and interpersonal relationships between its members. This not only takes into account the mutually negative relationships between its members, but also the combination of their capabilities to generate and resolve conflicts.